\documentclass[aps,pra,twocolumn,superscriptaddress,showpacs, amsmath, amssymb]{revtex4-1}
\usepackage{graphicx}
\usepackage{bm}% bold math
\usepackage{amsfonts}
\usepackage{latexsym}

\usepackage[usenames]{color}
\usepackage{ulem}
\begin{document}
%%%%%%%%%%%%%%%%%%%%%%%%%%%%%%%%%%%%%%%%%%%%%%%%%%%%%%%
%Title
\title{Strong coupling between a trapped single atom and an all-fiber cavity}
%Authors
\author{Shinya Kato}
\affiliation{Department of Applied Physics, Waseda University, 3-4-1 Okubo, Shinjuku, Tokyo 169-8555, Japan}
\author{Takao Aoki}
\email{takao@waseda.jp}
\affiliation{Department of Applied Physics, Waseda University, 3-4-1 Okubo, Shinjuku, Tokyo 169-8555, Japan}

\begin{abstract}
We demonstrate an all-fiber cavity quantum electrodynamics system with a trapped single atom in the strong coupling regime.
We use a nanofiber Fabry--P\'{e}rot cavity, that is, an optical nanofiber sandwiched by two fiber-Bragg-grating mirrors.
Measurements of the cavity transmission spectrum with a single atom in a state-insensitive nanofiber trap clearly reveal the vacuum Rabi splitting.
\end{abstract}


\maketitle

%%%%%%%%%%%%%%%%%%%%%%%%%%%%%%%%%%%%%%%%%%%%%%%%%%%%%%%
%Introduction

Cavity quantum electrodynamics (QED) in the strong coupling regime, where the atom--cavity coupling rate exceeds all dissipation rates of the system, has been at the forefront of the exploration of the coherent dynamics of open quantum systems\cite{Miller2005,Haroche2013}.
For experimental studies at optical frequencies, remarkable progress has been made with single atoms trapped in free-space Fabry--P\'{e}rot cavities, ranging from observations of the vacuum Rabi splitting\cite{Boca2004}, photon blockade\cite{Birnbaum2005}, and squeezed light\cite{Ourjoumtsev2011} to realizations of a one-atom laser\cite{McKeever2003}, a deterministic single-photon source\cite{McKeever2004}, nondestructive detection of a photon\cite{Reiserer2013}, and a quantum gate between a photon and an atom\cite{Reiserer2014}.
Furthermore, the unique capabilities of cavity QED systems for storing and controlling the quantum states of atoms and light make them ideal nodes for a quantum network\cite{Kimble2008}, which has a wide variety of applications from the implementation of quantum computation\cite{Ladd2010} to fundamental studies on quantum many-body systems\cite{Georgescu2014}.
Quantum nodes are required to be capable of storing and controlling local quantum information as well as to be efficiently interfaced with the quantum channels through which flying quantum information is transmitted\cite{Northup2014}.
An elementary quantum network of two cavity QED systems has been demonstrated\cite{Ritter2012}.
Toward the realization of a large-scale quantum network, fiber-based alternatives are demanded to overcome the poor scalability of free-space Fabry--P\'{e}rot cavities. Recent advancements in the application of fiber-coupled photonic devices to the study of quantum optics have primarily focused on the Purcell (fast-cavity) regime of cavity QED, where the cavity dissipation dominates the system dynamics, or on the coupling between atoms and waveguides.
For example, coupling a trapped atom to a photonic-crystal cavity with a large cooperativity has been realized\cite{Thompson2013}, and a quantum phase switch has been demonstrated\cite{Tiecke2014}.
Efficient coupling between atoms and photonic-crystal waveguides has been also demonstrated\cite{Goban2014}.
Fiber-coupled whispering-gallery-mode (WGM) microcavities have been used to demonstrate various routing/switching schemes of photons in the Purcell regime\cite{Dayan2008, Aoki2009, OShea2013, Volz2014, Shomroni2014}. 
Although strong coupling between free-falling atoms and these WGM microcavities has been observed\cite{Aoki2006, Alton2011, Junge2013}, trapping an atom in the evanescent field of the WGMs still remains a challenge.

Here, we present the observation of strong coupling between trapped single cesium atoms and an all-fiber cavity.
Our cavity relies on tight transversal-mode confinement and the large evanescent fields of a nanofiber, which lead to efficient coupling of an atom and light, even with a single pass of the propagating guided mode, as intensively studied recently\cite{Nayak2007, Vetsch2010, Goban2012, Beguin2014}.
Therefore, strong atom--cavity coupling can be achieved with a relatively low cavity finesse and long cavity length\cite{Kien2009}.
Specifically, both ends of the nanofiber are connected through tapered regions to standard single-mode optical fibers with fiber-Bragg-grating (FBG) mirrors, thereby forming an all-fiber Fabry--P\'{e}rot cavity\cite{Wuttke2012} with a cavity finesse less than 40 and a cavity length of 33~cm.
By designing one of the FBGs to have its reflection-band edge at the resonance of cesium, the output coupling condition can be temperature-tuned from undercoupling to overcoupling.
Clear vacuum Rabi splitting is observed in the transmission spectrum of the cavity with an atom trapped in a state-insensitive nanofiber trap\cite{Kien2005, Lacroute2012, Goban2012}.
Our system paves the way toward the realization of a large-scale all-fiber quantum network.

%%%%%%%%%%%%%%%%%%%%%%%%%%%%%%%%%%%%%%%%%%%%%%%%%%%%%%%%
%Figure 1
\begin{figure*}[htbp]
\begin{center}
   \includegraphics[width=\textwidth]{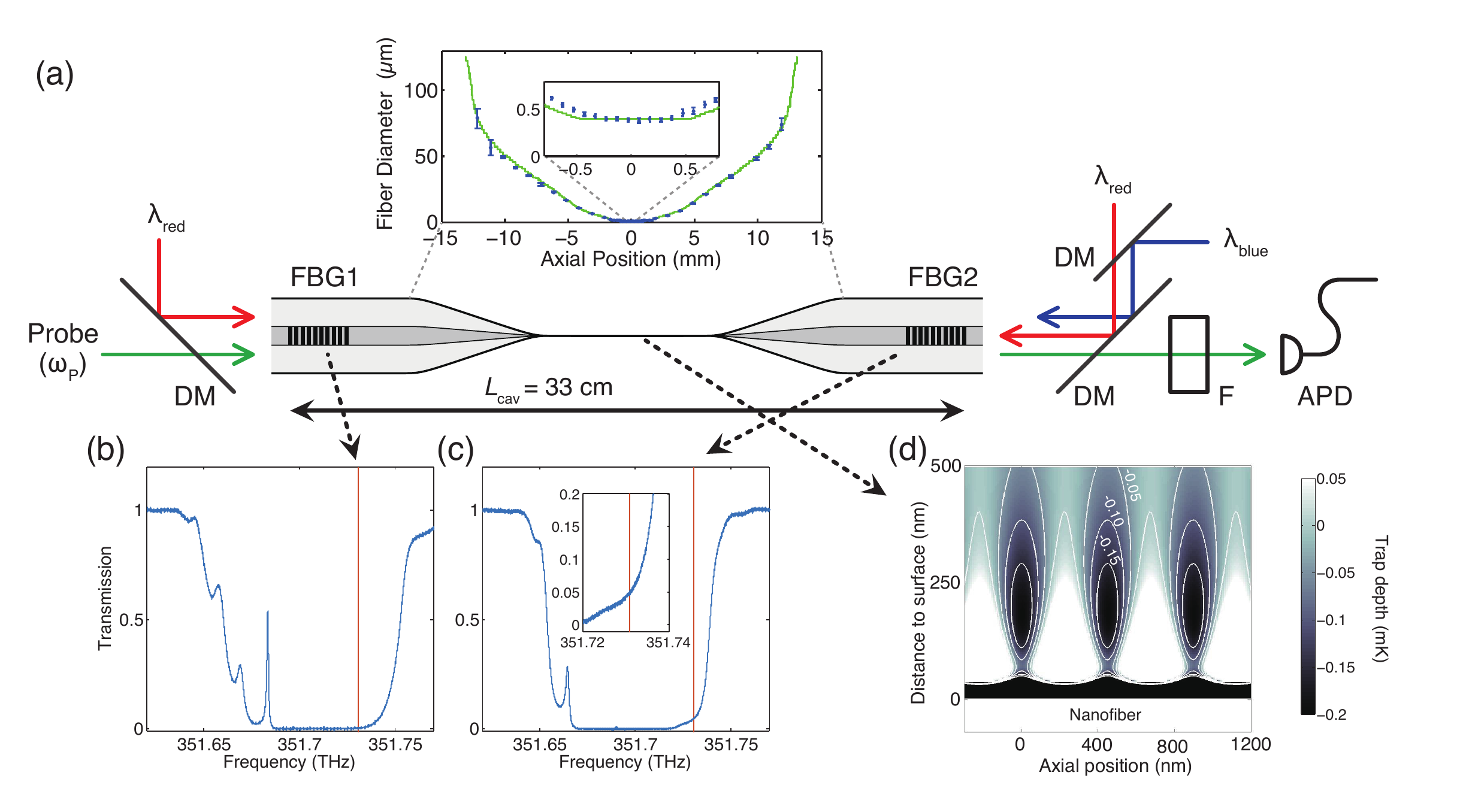}
\caption{
(a) Simple schematic of the experiment. Two FBG mirrors form a Fabry--P\'{e}rot cavity with a nanofiber waist.
Two-color trapping beams ($\lambda_{\rm blue}, \lambda_{\rm red}$) create a nanofiber trap for an atom.
A probe pulse with a frequency $\omega_{\rm P}$ is used to detect the presence of an atom and to measure the vacuum Rabi spectra.
DM: dichroic mirror. F: filters. APD: avalanche photodetector.
The inset shows the profiles of the nanofiber and tapered regions measured by a scanning electron microscope (blue circles) and a numerical model (green solid line).
(b), (c) Transmission spectra separately measured for FBG1 and FBG2, respectively.
The red solid line indicates the atomic resonance frequency, $\omega_{\rm A}$ ($6^2{\rm S}_{1/2}; F=4 \rightarrow 6^2{\rm P}_{3/2}; F^\prime = 5^\prime$ transition of cesium).
The inset in (c) is a magnified plot for the region around $\omega_A$.
(d) Trap potential in the vicinity of the nanofiber surface.
The trap frequencies in the axial, radial, and azimuthal directions are $2\pi \times (267, 159, 36)$ kHz, respectively.
Details of the trap potential structure can be found in Ref.~\cite{Lacroute2012}.
}
\label{fig-1}
\end{center}
\end{figure*}

%%%%%%%%%%%%%%%%%%%%%%%%%%%%%%%%%%%%%%%%%%%%%%%%%%%%%%%
%Experimental setup
Figure~\ref{fig-1} shows a schematic of the experiment.
A nanofiber is fabricated as the waist of the tapered optical fiber by using a homemade fiber-pulling rig described in Ref.~\cite{Nagai2014}.
The pulling sequence is numerically optimized to suppress the transmission losses through the tapered regions, and transmission exceeding 99~\% is routinely obtained.
A typical fabrication result and the numerically optimized shape are shown in the inset of Fig.~\ref{fig-1}(a).
The diameter and length of the nanofiber region are 400~nm and 1~mm, respectively.
Two FBG mirrors for the high reflector (FBG1) and output coupler (FBG2) form a one-sided Fabry--P\'{e}rot cavity.
The cavity length is estimated to be $L_{\rm cav} =$ 33~cm from the measurement of the free spectral range.
The total cavity field decay rate is given by $\kappa = \kappa_1 + \kappa_2 + \kappa_{\rm loss}$, where $\kappa_i$ and $\kappa_{\rm loss}$ are the field decay rate through FBG$i$ and the intracavity losses per roundtrip, respectively.
Figures~\ref{fig-1}(b) and (c) show the transmission spectra of FBG1 and FBG2, respectively, measured separately before the construction of the cavity.
The reflectivity of FBG1 is 99.5~\%; hence, $\kappa_1 = 2\pi\times$ 0.12\ MHz.
We design the output coupler, FBG2, to have its reflection-band edge at the D$_2$-line ($6^2{\rm S}_{1/2} \rightarrow 6^2{\rm P}_{3/2}$) transition of cesium, as can be clearly seen in Fig.~\ref{fig-1}(c).
Using the temperature-dependent shift in the reflection spectrum, the reflectivity of FBG2 (hence, $\kappa_2$) can be tuned by its temperature (Appendix).
The atom-trapping scheme is based on the pioneering works on the nanofiber trap in Refs.~\cite{Kien2004, Kien2005, Vetsch2010, Lacroute2012, Goban2012}.
Counter-propagating red-detuned ($\lambda_{\rm red}=937$~nm) trapping beams and a blue-detuned ($\lambda_{\rm blue}=688$~nm) trapping beam are input into the fiber by reflecting them off dichroic mirrors.
These wavelengths are chosen to be the so-called magic wavelengths, where the state-dependent scalar light shifts are canceled for the D2-line transition of cesium atoms\cite{Kien2005, Lacroute2012, Goban2012}.
Note that these wavelengths are outside of the reflection bands of the FBGs, and the propagation of the trapping beams is not affected by the FBGs.
Therefore, the trapping potential is created in the same manner as the nanofiber traps without FBGs\cite{Vetsch2010,Goban2012,Beguin2014}.
The transmission of a probe pulse with a frequency $\omega_{\rm P}$ is detected by an avalanche photodetector after blocking unwanted stray light using filters.

%%%%%%%%%%%%%%%%%%%%%%%%%%%%%%%%%%%%%%%%%%%%%%%%%%%%%%%%
%Cavity charecteristics
We first characterize the empty cavity (in the absence of an atom) at various temperatures (Appendix).
From the photon lifetime obtained in the cavity ring-down measurement with the critical-coupling condition ($\kappa_2 = \kappa_1 + \kappa_{\rm loss}$), which occurs at temperature $T=22.6\,^\circ {\rm C}$, we estimate $\kappa_{\rm loss}=2\pi\times3.2$~MHz, corresponding to the one-way transmission of the tapered optical fiber of 94~\%.
The degradation in the transmission of the tapered optical fiber is presumably due to contamination of the nanofiber region during the installation of the cavity into the vacuum chamber.

%%%%%%%%%%%%%%%%%%%%%%%%%%%%%%%%%%%%%%%%%%%%%%%%%%%%%%%%
%Figure 2 (Original: Fig. 3)
\begin{figure}[htbp]
\begin{center}
   \includegraphics[width=0.5\textwidth]{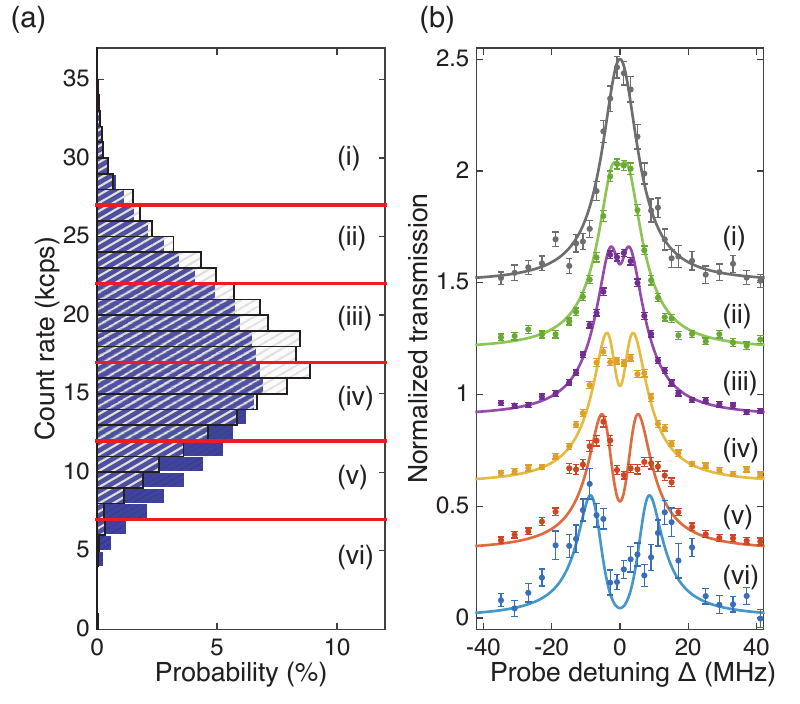}
\caption{
(a) Histogram of the transmission intensity of the detection probe with (blue) and without (gray) the optical molasses.
The dark count of the detector is 1.4$\times 10^3$  counts per second (cps), and the measurement back ground noise including the dark count is about 10 kcps.
(b) Transmission spectra as functions of the probe detuning $\Delta$.
Data sets and fits are normalized to the empty-cavity transmission and are vertically offset for clarity.
The observed asymmetry in the spectra is presumably due to the effect of the probe pulse on the center-of-mass motion of the atom.
Error bars are the standard error of the mean.
The atom--cavity coupling rates $g$ for the fits are $2\pi\times(1.3, 1.9, 2.9, 4.3, 7.8)$~MHz for (ii)--(vi), respectively.
}
\label{fig-3}
\end{center}
\end{figure}

%%%%%%%%%%%%%%%%%%%%%%%%%%%%%%%%%%%%%%%%%%%%%%%%%%%%%%%%
%Experiment 1
We next measure the transmission spectra of the atom--cavity coupled system.
The temperature of the cavity is set to have a critical coupling condition ($T\approx 22.6\, ^\circ {\rm C}$), and the cavity resonance frequency is set within $\pm 10$~MHz of the atomic resonance, $\omega_{\rm A}$.
We use a pair of counter-propagating red-detuned ($\lambda_{\rm red}=937$~nm) beams, each with a power of 0.2~mW, and a blue-detuned ($\lambda_{\rm blue}=688$~nm) beam with a power of 3.4~mW for the nanofiber trap\cite{Vetsch2010, Goban2012, Beguin2014}.
The wavelengths of the trapping beams are chosen as the magic wavelengths, where the differential scalar light shifts are eliminated for the D2-line transition of cesium atoms\cite{Kien2005, Lacroute2012, Goban2012}.
Note that we do not use counter-propagating beams for the blue-detuned trapping field because of technical reasons.
Therefore, differential vector light shifts are not suppressed, unlike the compensated nanofiber trap demonstrated in Ref.~\cite{Goban2012}.
The polarizations of the trapping fields are linearly polarized and parallel to each other.
The optical trap depth is set to 210~$\mu$K, and the potential minimum is located approximately 170~nm from the nanofiber surface.

Each measurement sequence starts by loading an atom into the nanofiber trap from a standard six-beam optical molasses, which
is spatially overlapped with the nanofiber region of the cavity to provide a cold and dilute cloud of cesium atoms with an atomic density as low as $2.5\times 10^5\ {\rm cm}^{-3}$.
To make the probability of loading multiple atoms into a trap negligible, we deliberately set the atomic density low, resulting in a low loading efficiency. 
We use the D$_2$-line $F=4 \rightarrow F^\prime=5^\prime$ transition for cooling and the $F=3 \rightarrow F^\prime=4^\prime$ transition for repumping in the optical molasses.
The detuning of the cooling beams is $-1.6 \Gamma$, and the total intensity is $8 I_{\rm s}$ in the loading stage, where $\Gamma$ and $I_{\rm s}$ are the natural linewidth and the saturation intensity of the cooling transition, respectively.
The loading time $\tau_{\rm load}$ is 30~ms.
After loading, we change the detuning and intensity to $-4.4 \Gamma$ and $3.7 I_{\rm s}$, respectively, and hold for 5~ms to allow for further cooling.

After the above molasses stage for atom loading and cooling, we send a resonant ($\omega_{\rm P} = \omega_{\rm A}$) probe pulse for detecting the presence of an atom (detection probe) with a power and pulse duration of 0.8~pW and 2~ms, respectively.
As shown in Fig.~\ref{fig-3}(a), we observe a small fraction of events with a lower transmission compared to the case without the molasses.
This lower transmission is the signature of the coupling between an atom and the cavity; the atom--cavity coupling shifts the resonance of the normal mode from $\omega_{\rm A}$ (vacuum Rabi splitting); hence, there is lower transmission at $\omega_{\rm A}$.
Note that the local atom--cavity coupling rate at the minimum of each potential well varies because the periods of the standing wave of the trap and the cavity mode are different.
The distribution of the atom--cavity coupling rate leads to the distribution of the reduction in transmission for the detection probe; a stronger atom--cavity coupling results in a larger reduction in transmission.
Therefore, we classify the reduction in detection-probe transmission into six levels ((i)--(vi) in Fig.~\ref{fig-3}(a)) and use this classification as a criterion for further investigation.

Following the detection pulse, we send another probe pulse with a variable detuning $\Delta = \omega_{\rm P} - \omega_{\rm A}$ for measuring the transmission spectra (spectroscopy probe) with a power and pulse duration of 0.4~pW and 5~ms, respectively\cite{Comment1}.
Figure~\ref{fig-3}(b) shows the observed spectra for each level of the reduction in detection-probe transmission (i)--(vi).
For case (i), the transmission spectrum exhibits a single Lorentzian, which indicates the absence of an atom.
From the Lorentzian fit to this spectrum, we obtain the total cavity field decay rate $\kappa=2\pi\times 6.4$~MHz, which is consistent with the photon lifetime of 12.5~ns, as measured with the cavity ring-down described above.
On the other hand, we observe broadening and splitting of the spectra for cases (ii)--(vi) because of atom--cavity coupling.

The steady-state transmission spectrum for the atom--cavity system in the weak-driving limit is given by\cite{Thompson1992}
\begin{eqnarray}
T(\Delta)=\left| \frac{2\sqrt{\kappa_1\kappa_2}(i\Delta+\gamma)}{(i\Delta+\kappa)(i\Delta+\gamma)+g^2} \right|^2,\label{eq:T}
\end{eqnarray}
where $g$ and $\gamma$ are the atom--cavity coupling rate and atomic polarization decay rate, respectively.
The atom--cavity coupling rate $g$ is given by\cite{Kimble1998}
\begin{eqnarray}
g({\bf r})=\sqrt{\frac{\mu^2 \omega}{2\hbar \epsilon_0 V_{\rm mode}}}\phi({\bf  r}),
\end{eqnarray}
where $\mu$ is the transition dipole moment, $\phi({\bf  r})$ is the cavity mode amplitude, and $V_{\rm mode}=\int|\phi({\bf  r})|^2 {\rm d}V$ is the cavity mode volume.
For calculating $V_{\rm mode}$, we neglect the contribution from the tapered regions and the nanofiber, and we use the fundamental mode (the hybrid HE$_{11}$ mode) of the single-mode fiber (SM800-5.6-125, Thorlabs, Inc.) at the wavelength of 852.3~nm:
\begin{eqnarray}
V_{\rm mode}=L_{\rm cav}\int_A|\phi_{{\rm HE_{11}}}|^2{\rm d}A,
\end{eqnarray}
where $A$ is the infinite cross section normal to the fiber axis. 
We obtain the numerically estimated value for the maximum coupling rate, $g_{\rm est}=2\pi\times 7.4$~MHz.

We fit Eq.~(\ref{eq:T}) to the observed transmission spectra for cases (ii)--(vi), where the only free parameter is $g$, and we obtain reasonable agreement, as shown in Fig.~\ref{fig-3}(b)(ii)--(vi). The measured maximum coupling constant $g_0=2\pi\times (7.8\pm1.2)$~MHz (Fig.~\ref{fig-3}(b)(vi)) agrees well with the numerically estimated value $g_{\rm est}=2\pi\times 7.4$~MHz.
Therefore, the strong coupling condition of $g_0 > (\kappa, \gamma)$ is achieved, where $(\kappa, \gamma) = 2\pi\times (6.4, 2.6)$~MHz.

%%%%%%%%%%%%%%%%%%%%%%%%%%%%%%%%%%%%%%%%%%%%%%%%%%%%%%%
%New expetiment 
To further confirm that only one atom is coupled to the cavity at a time, we measure the transmission spectra for various atom-loading times, $\tau_{\rm load}$\cite{Comment2}.
Figures~3(a)--(d) show the spectra for the events with the largest reduction level of detection-probe transmission (similar to Fig.~2(b)(vi)) for $\tau_{\rm load}=20, 10, 5, 2$ ms, respectively (Appendix).
As the loading time $\tau_{\rm load}$ is reduced, the probability of these events $P_{\rm (vi)}$ decreases; however, the coupling rate $g_0$ obtained from the fit to the observed spectra does not appreciably change.
This proves that the observed splitting indeed originates from a single atom coupled to the cavity.
%%%%%%%%%%%%%%%%%%%%%%%%%%%%%%%%%%%%%%%%%%%%%%%%%%%%%%%%
%New Figure 3
\begin{figure}[htbp]
\begin{center}
   \includegraphics[width=0.45\textwidth]{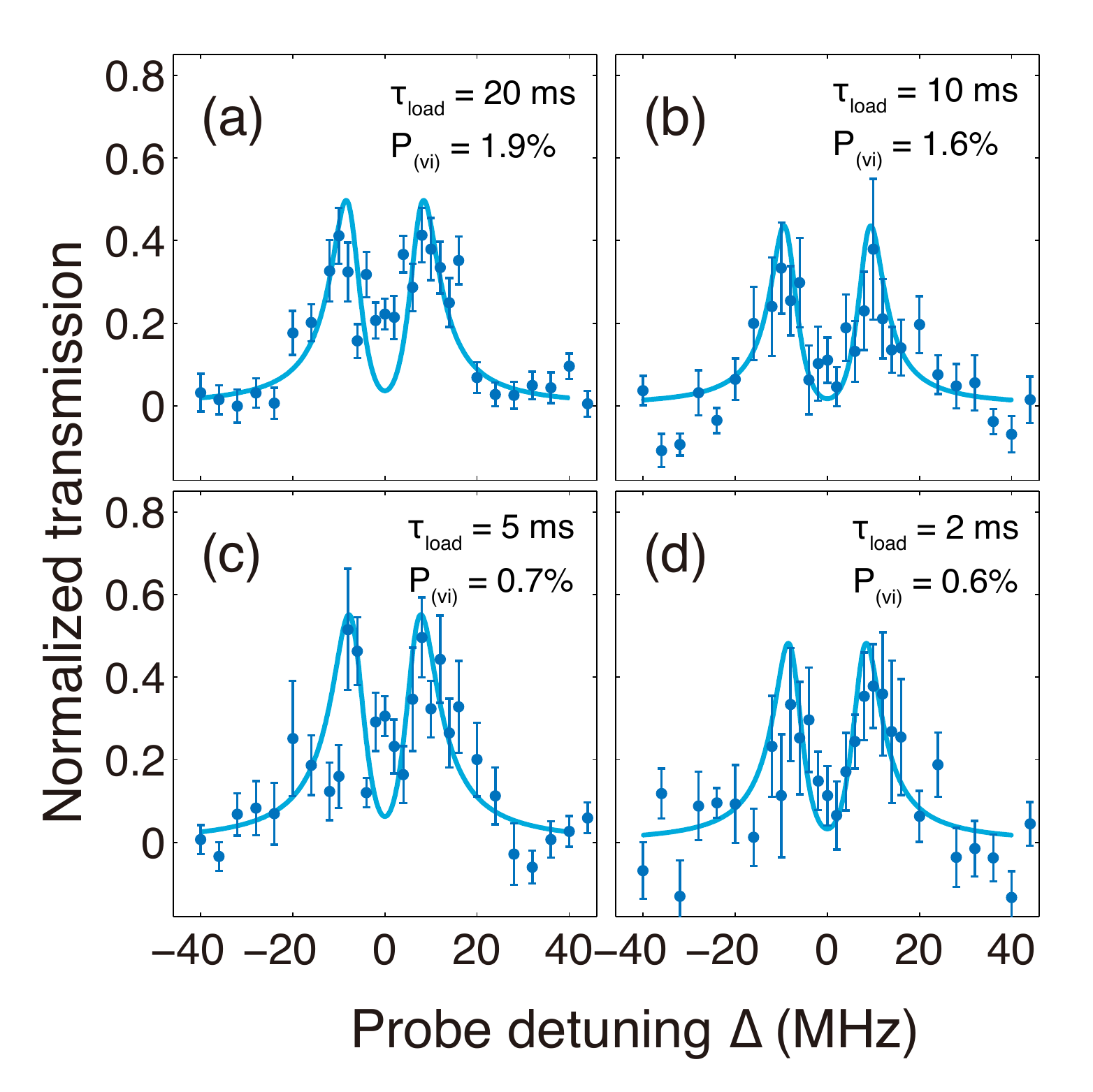}
\caption{
(a)--(d) Vacuum Rabi splitting spectra with various atom-loading times $\tau_{\rm load} = $20, 10, 5, and 2 ms, respectively.
The atom-cavity coupling rate $g$ determined from the fits are $2\pi\times(7.7\pm1.1, 8.9\pm1.4, 7.0\pm1.1, 7.8\pm1.1)$ MHz for (a)-(d), respectively.
The probability of the corresponding events $P_{\rm (vi)}$ is shown in each panel (Appendix).
}
\label{nfig-3}
\end{center}
\end{figure}

%%%%%%%%%%%%%%%%%%%%%%%%%%%%%%%%%%%%%%%%%%%%%%%%%%%%%%%%
%Figure 4
\begin{figure}[htbp]
\begin{center}
   \includegraphics[width=0.45\textwidth]{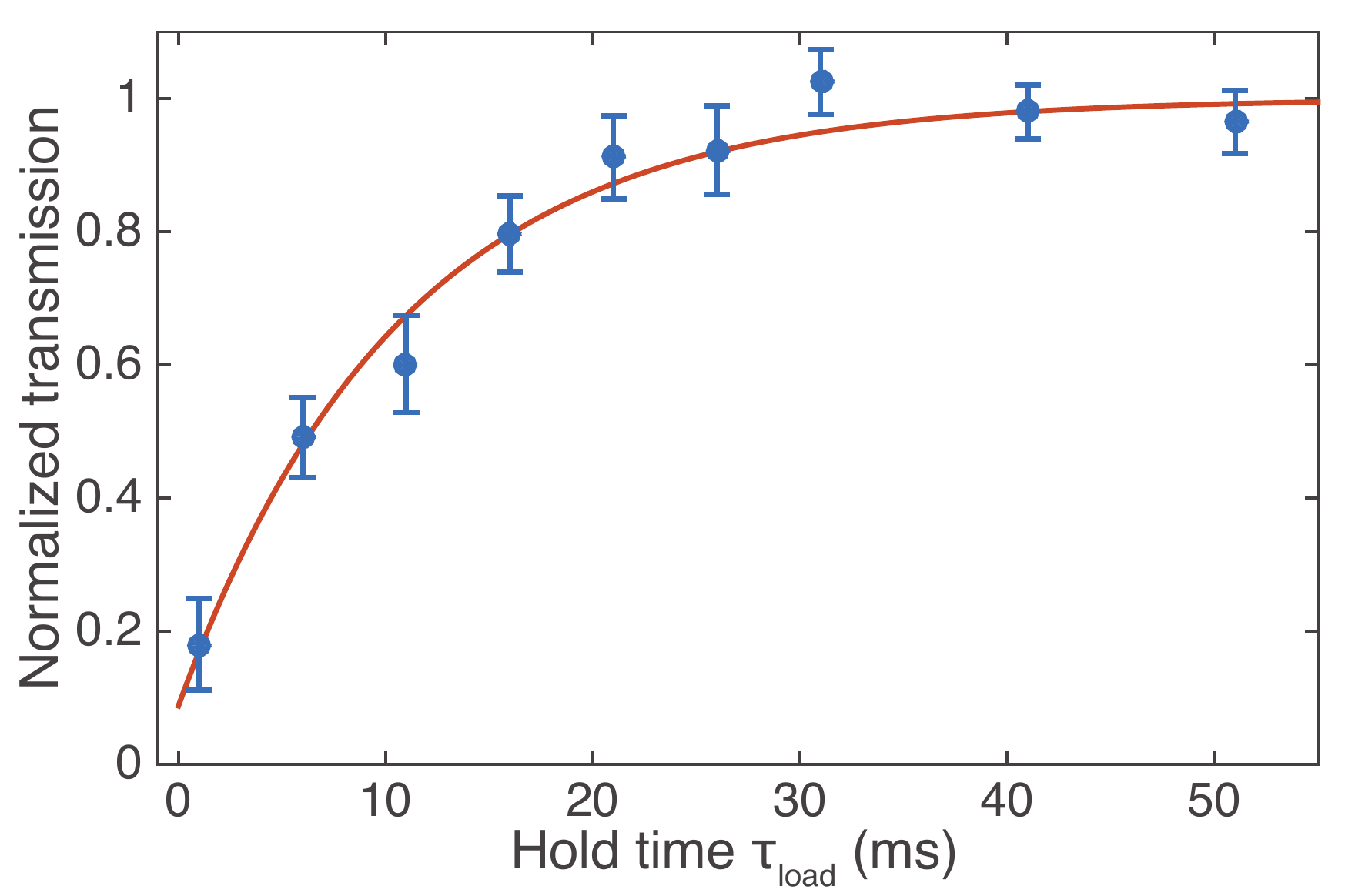}
\caption{
Normalized transmission as a function of the hold time.
The solid line is an exponential fit to the data.
Error bars denote the standard error of the mean.
}
\label{fig-4}
\end{center}
\end{figure}

%%%%%%%%%%%%%%%%%%%%%%%%%%%%%%%%%%%%%%%%%%%%%%%%%%%%%%%%
%Experiment 2
Lastly, we measure the lifetime of the atom trap.
We fix the detuning of the spectroscopy probe to be zero, {\it i.e.}, $\Delta = 0$.
We insert a variable hold time $\tau_{\rm hold}$ between the detection and spectroscopy probes, and we record the change in transmission as a function of the hold time.
As shown in Fig.~\ref{fig-4}, the transmission recovers for a longer hold time.
An exponential fit to the data gives a lifetime of 11~ms, which agrees with that observed in the nanofiber trap of $12\pm 1$~ms\cite{Goban2012}.
This lifetime is comparable to the time required for each measurement sequence for Figs.~2 and 3.
This means that there are some events of an atom escaping from the trap during the measurement sequence for Figs.~2 and 3, 
which may result in a slight underestimate of the atom--cavity coupling rate $g_0$ in the above analysis.
Note that the recoil-limited lifetime can be as long as tens of seconds.
The observed lifetime in our setup is probably limited by heating due to the intensity and polarization fluctuations in the trapping beams.

%%%%%%%%%%%%%%%%%%%%%%%%%%%%%%%%%%%%%%%%%%%%%%%%%%%%%%%%
%Summary

In summary, we have demonstrated strong coupling between trapped single atoms and an all-fiber cavity.
By combining an ultralow-loss tapered optical fiber with transmission $> 99.95 \%$\cite{Hoffman2014} and FBGs with reflectivity $> 99.9 \%$, a cooperativity $C=g_0^2/(2\kappa \gamma)> 150$ is within reach.
In addition to applications to all-fiber quantum networks, our nanofiber Fabry--P\'{e}rot cavity provides new avenues in quantum optics.
By loading from a dense magneto-optical trap\cite{Vetsch2010, Goban2012, Beguin2014}, more than a thousand atoms can be trapped in the nanofiber trap in the Lamb--Dicke regime with each of the atoms being strongly coupled to the cavity.
Because the free spectral range of our cavity is on the order of 100~MHz, it is possible to match its integral multiple with the hyperfine splittings of the ground or excited states of alkali atoms, realizing simultaneous coupling of the two transitions in a $\Lambda$- or V-type three-level system, both in the strong coupling regime.

\begin{acknowledgments}
We thank R. Nagai for discussions and assistance on taper fabrication and M. Iura for assistance during the preliminary stage of the construction of the setup.
This work was supported by JSPS KAKENHI Grant Numbers 26707022, 14J02311, MATSUO FOUNDATION, Research Foundation for Opto-Science and Technology, and Waseda University Grant for Special Research Projects (Project number:2013A-501).
\end{acknowledgments}

\appendix

\section{Cavity ring-down in the reflection geometry}

%%%%%%%%%%%%%%%%%%%%%%%%%%%%%%%%%%%%%%%%%%%%%%%%%%%%%%%%
%SI Fig 1
\begin{figure}[htbp]
\begin{center}
   \includegraphics[width = 0.5\textwidth]{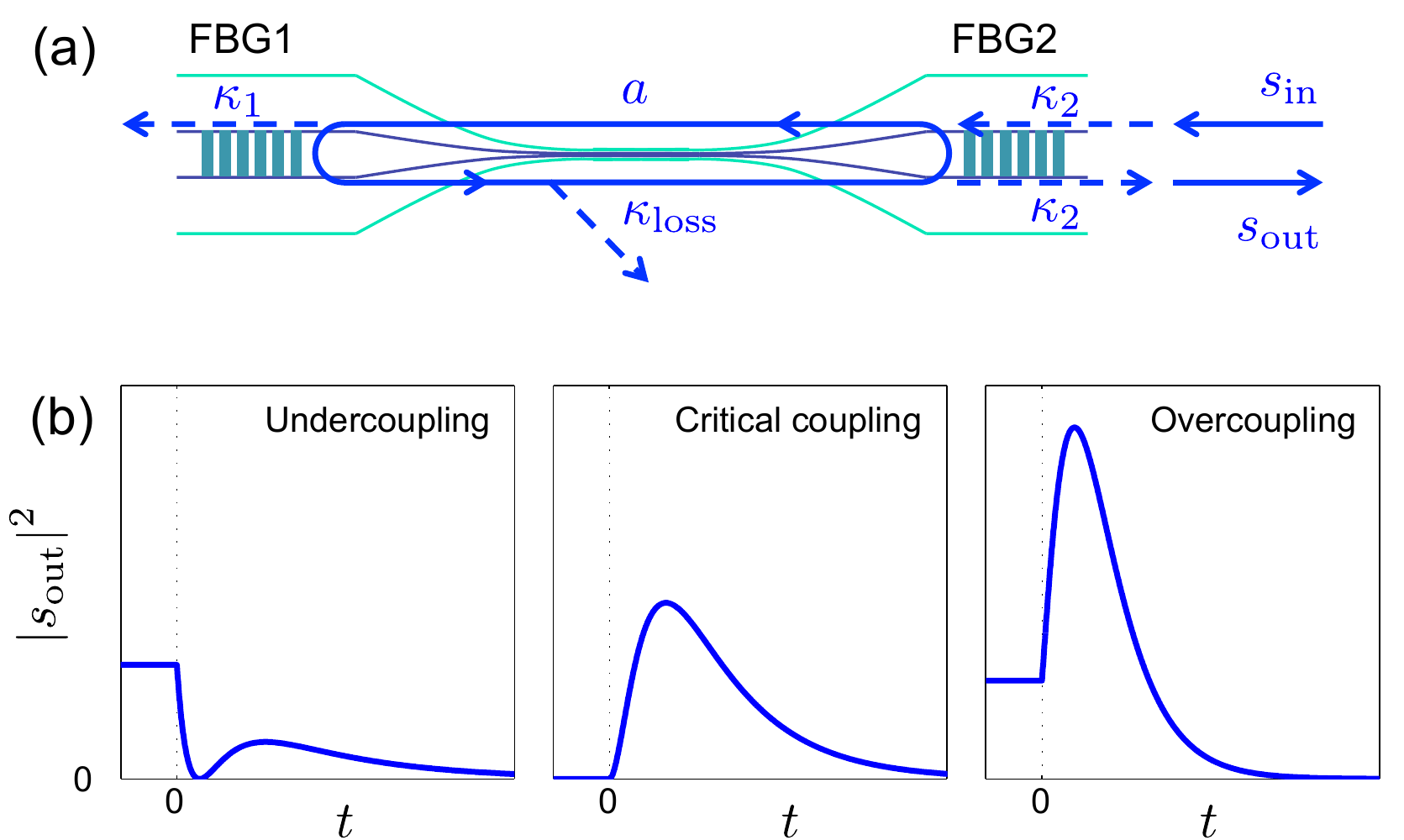}
\caption{
(a) Schematic of the cavity.
(b) Temporal profiles of the reflected intensity $|s_{\rm out}(t)|^2$ for undercoupling (left), critical coupling (middle), and overcoupling (right).
}
\label{si_fig-1}
\end{center}
\end{figure}

We describe a model for cavity ring-down in the reflection geometry. 
As shown in Fig.~\ref{si_fig-1}(a), two FBG mirrors for the high reflector (FBG1) and output coupler (FBG2) form a one-sided Fabry--P\'{e}rot cavity.
The total cavity field decay rate is given by $\kappa = \kappa_1 + \kappa_2 + \kappa_{\rm loss}$, where $\kappa_i$ and $\kappa_{\rm loss}$ are the field decay rate through FBG$i$ and the intracavity losses per roundtrip, respectively.
A resonant beam is input from the FBG2 port, and the intensity of the reflected beam is measured.
The temporal evolution of the cavity field amplitude $a$ is described by\cite{Haus1984}
\begin{eqnarray}
\frac{da}{dt} = i\omega_0 a -\kappa a + \sqrt{2\kappa_2} s_{\rm in},
\label{eq:EQM}
\end{eqnarray}
where $s_{\rm in}$ is the input field amplitude.
The reflected field amplitude $s_{\rm out}$ is given by 
\begin{eqnarray}
s_{\rm out} = -s_{\rm in} + \sqrt{2\kappa_2} a.
\label{eq:ref-field}
\end{eqnarray}
Note that $a$ is normalized to the energy, whereas $s_{\rm in, out}$ is normalized to the power.
We assume that the system is in the steady state, that the system is driven with a resonant input field with a constant amplitude for $t<0$, and that the input field is switched off at $t=0$ with a decay rate of $\kappa_{\rm s} \, (>\kappa)$,
\begin{eqnarray}
s_{\rm in} (t) = \left\{
\begin{array}{ll}
s_0e^{i\omega_0 t} & (t<0) \\
s_0e^{i\omega_0 t}e^{-\kappa_{\rm s}t} & (t \ge 0)
\end{array}
\right. .
\label{eq:in-field}
\end{eqnarray}
The substitution of  Eq.~(\ref{eq:in-field}) into Eq.~(\ref{eq:EQM}) results in
\begin{widetext}
\begin{eqnarray}
a (t) = 
\left\{
\begin{array}{ll}
\frac{\sqrt{2\kappa_2}}{\kappa} s_0 e^{i\omega_0 t} & (t<0) \\
\frac{\sqrt{2\kappa_2}}{\kappa} s_0 
\left[
\left(1 + \frac{\kappa}{\kappa_{\rm s}-\kappa} \right)e^{-\kappa t} - \frac{\kappa}{\kappa_{\rm s}-\kappa} e^{- \kappa_{\rm s}t}
\right]
e^{i\omega_0 t}
& (t \ge 0)
\end{array}
\right. ,
\label{eq:cav-field}
\end{eqnarray}
\end{widetext}
from which we obtain the reflected intensity,
\begin{widetext}
\begin{eqnarray}
|s_{\rm out}(t)|^2 = 
\left\{
\begin{array}{ll}
\left|  \frac{2\kappa_2}{\kappa} -1 \right|^2s_0^2 & (t<0) \\
\left| 
\left( \frac{2\kappa_2}{\kappa} +  \frac{2\kappa_2}{\kappa_{\rm s}-\kappa} \right)e^{-\kappa t}
-\left( 1 + \frac{2\kappa_2}{\kappa_{\rm s}-\kappa} \right) e^{-\kappa_{\rm s}t} 
\right|^2 s_0^2
& (t \ge 0)
\end{array}
\right. .
\label{eq:ref-int}
\end{eqnarray}
\end{widetext}
The reflected intensity $|s_{\rm out}(t)|^2$ exhibits different temporal profiles for undercoupling ($\kappa_2 < \kappa_1 + \kappa_{\rm loss}$), critical coupling ($\kappa_2 = \kappa_1 + \kappa_{\rm loss}$), and overcoupling ($\kappa_2 > \kappa_1 + \kappa_{\rm loss}$), as shown in Fig.~\ref{si_fig-1}(b).
Note that, for all cases, $|s_{\rm out}(t)|^2$ decays with a decay rate of $2\kappa$ at $t \gg (\kappa_s - \kappa)^{-1}$.

\section{Temperature-controlled cavity couplings}

\begin{figure}[htbp]
\begin{center}
   \includegraphics[width=0.4\textwidth]{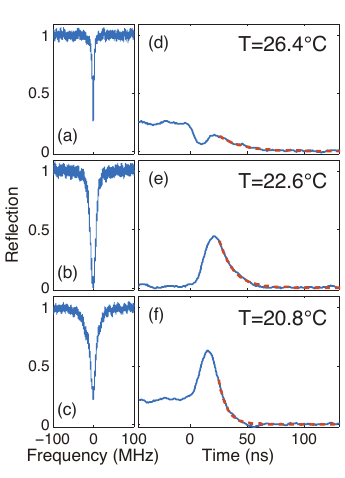}
\caption{
(a), (b), (c) Reflection spectra of the cavity from the FBG2 port at temperatures $T=26.4, 22.6,$ and $20.8\,^\circ {\rm C}$, respectively.
(d), (e), (f) Results of the cavity ring-down measurements for (a), (b), and (c), respectively; the dashed lines are exponential fits to the data with the constants of 18.4, 12.5, and 7.3~ns, respectively.
}
\label{nsi_fig-2}
\end{center}
\end{figure}

We characterize the empty cavity (in the absence of an atom) at various temperatures.
Figures~\ref{nsi_fig-2}(a), (b), and (c) show the reflection spectra of the cavity from the FBG2 port at temperatures $T=26.4, 22.6,$ and $20.8\,^\circ {\rm C}$, respectively.
Broadening of the cavity resonance linewidth at lower temperatures is clearly observed, which indicates an increase in $\kappa_2$.
This increase in $\kappa_2$ is more unambiguously confirmed by the cavity ring-down measurements, as shown in Figs.~\ref{nsi_fig-2}(d), (e), and (f).
The temporal profile of the ring-down reflection shows the change in the output coupling from undercoupling ($\kappa_2 < \kappa_1 + \kappa_{\rm loss}$) for (d) to critical coupling ($\kappa_2 = \kappa_1 + \kappa_{\rm loss}$) for (e) and overcoupling ($\kappa_2 > \kappa_1 + \kappa_{\rm loss}$) for (f), as described above.
Exponential fits to the tails of these traces gives the photon lifetimes ($(2\kappa )^{-1}$) of 18.4, 12.5, and 7.3~ns.

%%%%%%%%%%%%%%%%%%%%%%%%%%%%%%%%%%%%%%%%%%%%%%%%%%%%%%%%
%Figure 2

\section{Transmission spectra of the atom--cavity coupled system with various atom-loading times}

\begin{figure}[htbp]
\begin{center}
   \includegraphics[width=0.5\textwidth]{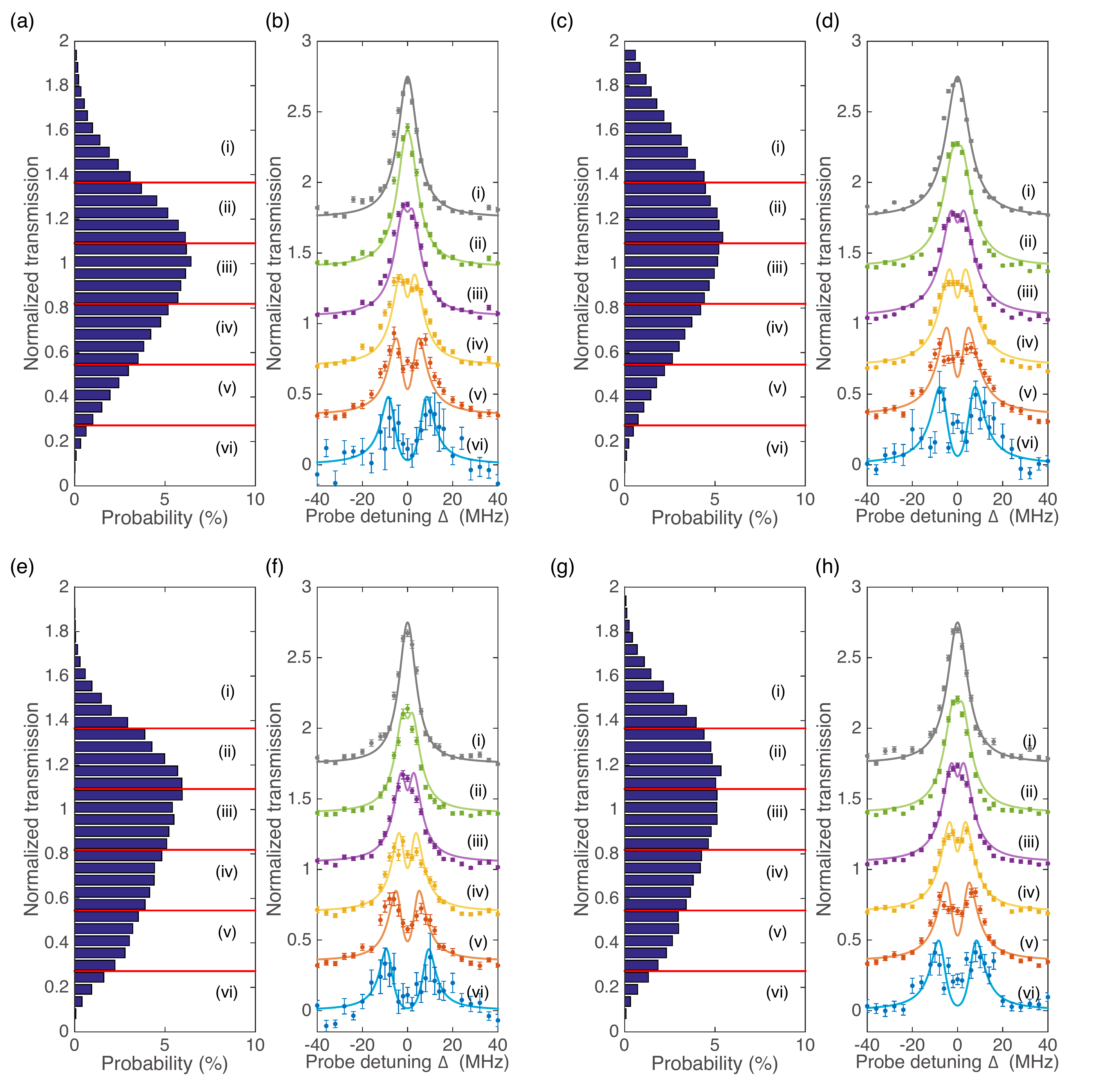}
\caption{
Spectroscopy results with various atom-loading times.
(a),(c),(e),(g) Histogram of the normalized intensities of the detection probe.
(b),(d),(f),(h) Transmission spectra as function of the probe detuning $\Delta$.
The loading times are 2 ms for (a) and (b), 5 ms for (c) and (d), 10 ms for (e) and (g), and 20 ms for (g) and (h).
}
\label{nsi_fig-3}
\end{center}
\end{figure}

Figure~\ref{nsi_fig-3} shows the transmission spectra of the atom--cavity coupled system with various atom-loading times.
For these measurements, the detuning and the total intensity of the molasses beams in the loading stage are $-4.4 \Gamma$ and $8 I_{\rm s}$, respectively, and there is no additional cooling stage after the loading.
The intensities of the detection and spectroscopy probes are 24 pW and 2.4 pW, respectively.
The durations of the detection and spectroscopy probe pulses are 2 ms and 5 ms, respectively.
To remove long-term drift, we normalized the detection-probe transmission by using the transmission signal of the frequency-scanning pulse for locating the cavity resonance frequency (see the footnote [36] in the main text).
In the same manner as Fig.~2 in the main text, we classify the reduction in the detection-probe transmission into six levels ((i)--(vi) in Fig.~\ref{nsi_fig-3}(a),(c),(e),(g)) and fit Eq.~(1) in the main text to each of the transmission spectra. 
The spectra for case (vi) and the corresponding values for $g_0$ are used in the main text.

%%%%%%%%%%%%%%%%%%%%%%%%%%%%%%%%%%%%%%%%%%%%%%%%%%%%%%%%
%Figure 3

%%%%%%%%%%%%%%%%%%%%%%%%%%%%%%%%%%%%%%%%%%%%%%%%%%%%%%%%

%%%%%%%%%%%%%%%%%%%%%%%%%%%%%%%%%%%%%%%%%%%%%%%%%%%%%%%%


\begin{thebibliography}{99}
\bibitem{Comment1}
After the detection and spectroscopy probe pulses, another pulse with a duration of 10~ms and a frequency scanned over several free spectral ranges is input, and its transmission is measured to locate the cavity resonance frequency.
The measured detuning of the cavity from the atomic $F=4 \rightarrow F^\prime=5^\prime$  transition is used as an error signal for locking the cavity frequency to the atomic transition.
In the analysis of the experimental data, we discard a sequence if the cavity detuning exceeds $\pm$10~MHz, which occasionally occurs owing to the instability of the lock.
\bibitem{Comment2}
For these measurements, the detuning and the total intensity of the molasses beams in the loading stage are $-4.4 \Gamma$ and $8 I_{\rm s}$, respectively, and there is no additional cooling stage after the loading.
The intensities of the detection and spectroscopy probes are 24 pW and 2.4 pW, respectively.
See Appendix for details.
\end{thebibliography}
\end{document}